\documentclass[doublecol]{epl2}

\title{Superconductivity at 25 K in hole doped $(La_{1-x}Sr_x)OFeAs$}

\author{Hai-Hu Wen*, Gang Mu, Lei Fang, Huan Yang, \and Xiyu Zhu}
\shortauthor{Hai-Hu Wen et al.}

\institute{National Laboratory for Superconductivity, Institute of
Physics and Beijing National Laboratory for Condensed Matter
Physics, Chinese Academy of Sciences, P. O. Box 603, Beijing 100080,
People's Republic of China}

\abstract{By partially substituting the tri-valence element La with
di-valence element Sr in $LaOFeAs$, we introduced holes into the
system. For the first time, we successfully synthesized the hole
doped new superconductors $(La_{1-x}Sr_x)OFeAs$. The maximum
superconducting transition temperature at about 25 K was observed at
a doping level of x = 0.13. It is evidenced by Hall effect
measurements that the conduction in this type of material is
dominated by hole-like charge carriers, rather than electron-like
ones. Together with the data of the electron doped system
$La(O_{1-x}F_x)FeAs$, a generic phase diagram is depicted and is
revealed to be similar to that of the cuprate superconductors. }

\begin{document}

\maketitle

\section{Introduction}

Superconductivity is a quantum phenomenon that shows the vanishing
of resistivity and exclusion of magnetic field due to the
condensation of paired electrons. A discovery of high temperature
superconductors not only brings about enormous scientific interests,
but also leads to potential applications. Besides the high
temperature superconductivity in the cuprate system that was firstly
found in 1986\cite{BednorzMuller}, and that in $MgB_2$ found in
2001\cite{MgB2}, efforts in exploring new materials lead to the
discovery of superconductivity in many other systems, such as
$Na_xCoO_2 \cdot 1.3H_2O$ \cite{NaCoO2}, $Sr_2RuO_4$\cite{SrRuO} and
etc., but all these have the transition temperatures below 20 K.
Searching new superconductors with 3d or 4d transition metal
compounds is specially interesting since the relatively strong
localization effects of electrons in these materials quite often
lead to strong correlation effects. In 1995, the fabrication of a
series of quaternary oxy-pnictides in a general formula as LnOMP
(where Ln= La-Nd, Sm and Gd; M = Mn, Fe, Co, Ni and Ru) was
published\cite{Zimmer}. The system has a layered structure and a
tetragonal P4/nmm space group, with a stacking series of
$-(LnO)_2-(MP)_2-(LnO)_2-$. In one unit cell, there are two
molecules of LnOMP, and it is valence self-balanced, i.e.,
(LaO)$^{+1}$ is balanced by (MP)$^{-1}$. Some of them, such as
LaOFeP and LaONiP, were shown to be superconductors at about 4
K\cite{Kamihara2} and 3 K\cite{Watanabe}, respectively. By
substituting the oxygen with F, the $T_c$ was increased to 7
K\cite{Kamihara2}. These iron based materials constructed a new
family of layered superconductors without copper. Very recently,
Kamihara et al.\cite{Kamihara1} found that by substituting P with
As, and by substituting partially the O in LaOFeP with F, the
resultant material $La(O_{1-x}F_x)FeAs$ (x = 0.05 to 0.12) became
superconductive at 26 K. This is really surprising since the iron
elements normally give rise to magnetic moments, and in many cases
they form a long range ferromagnetic order, and are thus detrimental
to the superconductivity with singlet pairing. This interesting
discovery has already attracted intense
efforts\cite{Singh,FangZ,Kotliar,MuG,ZhuXY} both from experimental
and theoretical side. Since the substitution of $O^{2-}$ by $F^{-}$
can introduce more electrons into $LaOFeAs$, it was called as
electron-doped. Interestingly, by substituting $La^{3+}$ with
$Ca^{2+}$ which brings more holes into the system, Kamihara et
al.\cite{Kamihara1} found no trace of superconductivity and
suggested that: a critical factor for induction of superconductivity
in this system is electron doping, and not hole doping. In this
Letter, we show the evidence of superconductivity in $LaOFeAs$
achieved by substituting $La^{3+}$ with $Sr^{2+}$, that is through
hole doping. The highest transition temperature found here is about
25 K.
\section{Sample preparation and experiment}

By using a two-step method, we successfully fabricated the
$(La_{1-x}Sr_x)OFeAs$ (x = 0.10 - 0.30) and
$La(O_{0.9}F_{0.1-\delta})FeAs$\cite{ZhuXY} samples. First the
starting materials Fe powder (purity 99.95\%) and As grains (purity
99.99\%) were mixed in 1:1 ratio, grounded and pressed into a pellet
shape. Then it was sealed in an evacuated quartz tube and followed
by burning at 700 $^\circ$C for 10 hours. The resultant pellet was
smashed and grounded together with the $SrCO_3$ powder (purity
99.9\%), $La_2O_3$ powder (purity 99.9\%) and grains of La (purity
99.99\%) in stoichiometry as the formula $(La_{1-x}Sr_x)OFeAs$.
Again it was pressed into a pellet and sealed in an evacuated quartz
tube and burned at about 900-940 $^\circ$C for 4 hours, followed by
a burning at 1150-1200 $^\circ$C for 48 hours. Then it was cooled
down slowly to room temperature. In Fig. 1(a), we show the X-ray
diffraction (XRD) patterns for the sample
$(La_{0.87}Sr_{0.13})OFeAs$. It is found that the peaks from XRD are
dominated by the phase of $LaOFeAs$ for low doping (below about
x=0.15) although some impurity peaks appear also. Beyond that
doping, some strong peaks from the impurity phase emerge and are
getting stronger with more doping. But the XRD taken from all
samples gives clear evidence that the main peaks are from the phase
$(La_{1-x}Sr_x)OFeAs$. From Fig. 1(a), it is clear that almost all
main peaks can be indexed by a tetragonal structure with $a = b =
4.0350 \AA$ and $c = 8.7710 \AA$. These lattice constants are a bit
larger than those in the parent phase $LaOFeAs$ ($a = b = 4.032 \AA$
and $c = 8.726 \AA$), suggesting that the lattice expands a bit with
Sr substitution, especially along the c-axis. This is understandable
since the radius of $Sr^{2+}$ is 1.12$\AA$ which is larger than that
of $La^{3+}$ (1.06 $\AA$). Therefore, it is certain that the
dominant component is from $(La_{0.87}Sr_{0.13})OFeAs$. There are
several peaks marked by the asterisks which may come from the
impurity phase of FeAs and LaAs. In Fig. 1(b) we present the energy
dispersive x-ray microanalysis (EDX) spectrum of one typical grain,
which shows that the main elements of the grains are La, Sr, Fe, As
and O. It is thus safe to conclude that the superconductivity
observed here comes from the main phase $(La_{1-x}Sr_x)OFeAs$.

\begin{figure}
\onefigure[width=8cm]{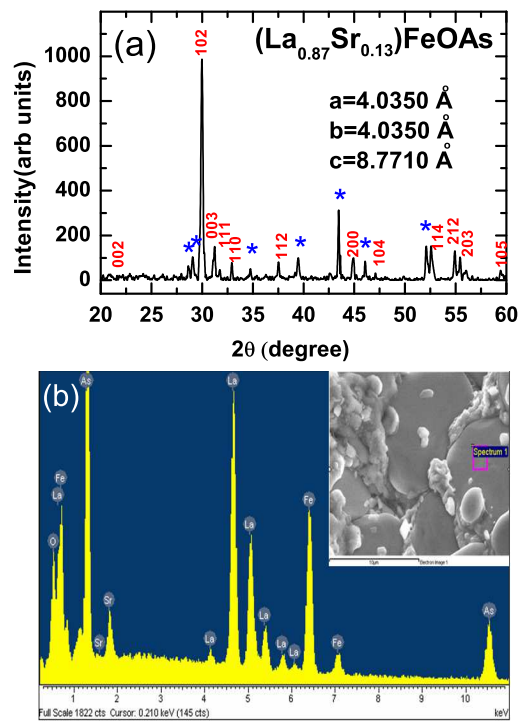} \caption{ (a) X-ray diffraction
patterns for the sample $(La_{0.87}Sr_{0.13})OFeAs$. All main peaks
can be indexed by a tetragonal structure with $a = b = 4.0350 \AA$
and $c = 8.7710 \AA$, indicating that the dominant phase here is
$(La_{0.87}Sr_{0.13})OFeAs$. The asterisks mark the peaks from the
impurity phase. (b) The EDX spectrum taken from one of the the main
grains, which shows that the main elements of the grains are La, Sr,
Fe, As and O. The inset in (b) shows a scanning electron microscopic
picture. The little rectangle marks the position where we took the
EDX spectrum. } \label{fig.1}
\end{figure}

The magnetic measurements were done with a superconducting quantum
interference device (Quantum Design, SQUID, MPMS7), and an Oxford
cryogenic system Maglab-12T. The AC susceptibility of the samples
were measured with the Maglab-12T with an AC field of 0.1 Oe and a
frequency of 333 Hz. The superconductivity was also proved by the DC
magnetization measurements using the zero-field-cooled mode, that is
by cooling the sample at zero field to 2 K, then applying a magnetic
field and the data was collected during the warming up process. The
resistivity and Hall effect measurements were done with a physical
property measurement system (Quantum Design, PPMS9T) with a
six-probe technique. The current direction was changed for measuring
each point in order to remove the contacting thermal power.

\section{Results and discussion}

\begin{figure}
\onefigure[width=8cm]{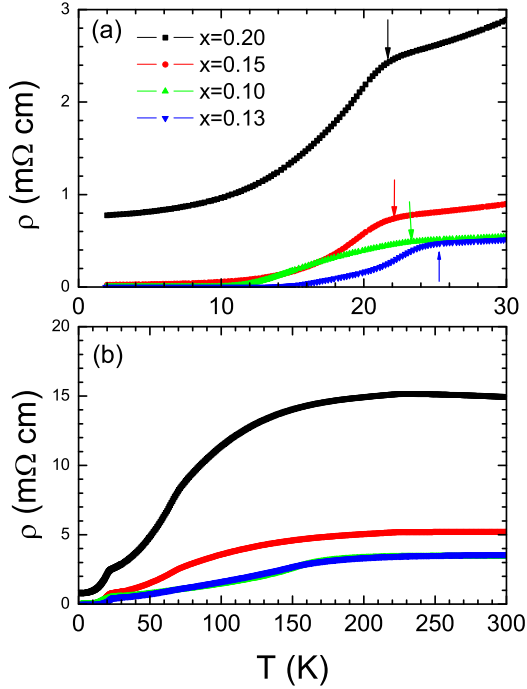} \caption{The temperature dependence
of resistivity of samples $(La_{1-x}Sr_x)OFeAs$ with the Sr
concentration x changing from 0.10 to 0.20. One can see that the
onset transition temperatures marked here by arrows are quite close
to each other, with the highest $T_c \approx 25.6 K$ at the doping
of 0.13. Beyond x = 0.20, no superconductivity was observed. }
\label{fig.2}
\end{figure}

\begin{figure}
\onefigure[width=8cm]{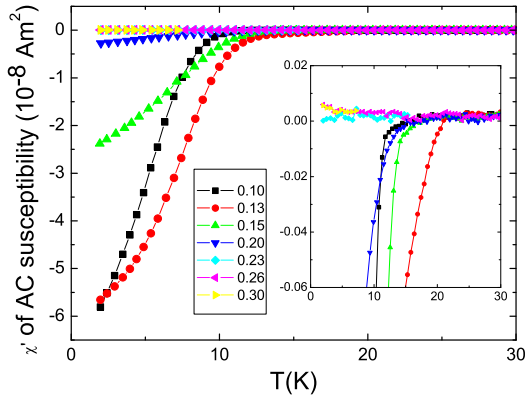} \caption{The temperature dependence
of the real part of the AC susceptibility measured with an AC field
of 0.1 Oe and the oscillating frequency of 333 Hz. The inset shows
an enlarged view of the same data in the main frame, with also the
same coordinates. The onset transition temperature behaves in the
same way as that determined from the resistivity data. }
\label{fig.3}
\end{figure}

In Fig.2 (a) we present the temperature dependence of the resistive
transitions for samples $(La_{1-x}Sr_x)OFeAs$ with x = 0.10 to 0.20.
One can see that the onset transition temperature taken with a
criterion of 95\% $\rho_n$ shifts slowly to higher values with the
doping amount of Sr from 0.10 to 0.13. The maximum onset $T_c
\approx$ 25.6 K is achieved at a doping of x = 0.13 and the zero
resistance temperature is about 15 K. Then onset $T_c$ drops down
slowly with further doping and becomes zero at the doping level of
about 0.23. This doping dependence is quite similar to that in
electron doped samples $La(O_{1-x}F_x)FeAs$ where the $T_c$ is
rather stable in the middle doping regime (x = 0.05 to 0.11). This
flattening behavior of $T_c$ is very different from that in cuprate
superconductors in which a parabolic doping dependence was observed.
At this moment, our samples are not pure enough, therefore the
transitions are still broad, and the $T_c$ values determined here
may change in the clean or pure samples. Thus the rather stable
$T_c$ versus doping may be an intrinsic effect of the material, or
it may be induced by the inhomogeneous phase formed during the
reaction. In Fig.2 (b) the resistivity in wide temperature region is
shown for the same samples. A huge bump appears for all samples in
high temperature region, which may reflect an unusual electron
scattering process or a drastic change of electron phonon
scattering. We will see later that, just corresponding to the
appearance of this bump, the Hall coefficient drops down quickly.
This huge bump appears for all samples here, which may have the same
origin as that appearing in the electron doped samples as marked as
$T_{anom}$ in \cite{Kamihara1}. By increasing the doping level, the
general resistivity gets larger. This is an interesting behavior,
again we are not sure whether this is an intrinsic effect of the
$(La_{1-x}Sr_x)OFeAs$ phase, or it is due to an extrinsic effect,
such as much stronger impurity scattering by more impurity centers
at a high doping level. This remark may be true since the sample x =
0.20 with larger normal state resistivity has not come into the
complete superconducting state even when the temperature goes down
to 2 K.

In Fig.3 we show the temperature dependence of the dia-magnetization
measured using AC susceptibility based on an Oxford cryogenic system
Maglab-12T. Although the transitions are still broad, an enlarged
view shown in the inset allow us to determine the onset magnetic
transition point. It is known that the magnetic onset transition
point is normally close to the resistive transition point at 50-90\%
$\rho_n$. Therefore the magnetic onset $T_c$ values determined in
this way are a bit lower than that determined from the onset
transition of resistivity. But it is clear that the onset transition
temperature determined on the magnetic signal follows the same way
as the resistive data. Both the resistive and magnetic transition
curves are still not perfectly sharp, which leaves more room for
improving the sample quality in the future work. But this does not
give any doubt about the superconducting transition temperatures
determined here. It is worthy to mention that, as reported in the
original paper for the electron doped samples\cite{Kamihara1}, a
magnetic background appears for all samples investigated here. The
magnetization-hysteresis-loop measurements above $T_c$ indicate that
it has a weak ferromagnetic feature. We don't know whether this
magnetic signal is an intrinsic property of the $LaOFeAs$ phase, or
it is due to the impurity phase. If the former case is true, the
superconductivity in the present system should be categorized as a
non-conventional one.

\begin{figure}
\onefigure[width=8cm]{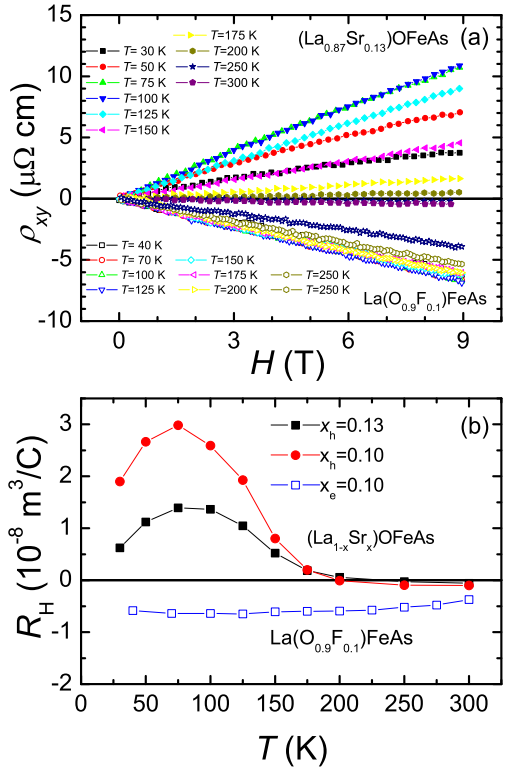} \caption{Hall effect measurements
for one sample $(La_{0.87}Sr_{0.13})OFeAs$ in present work and an
electron doped sample $La(O_{0.9}F_{0.1})FeAs$. (a) Magnetic field
dependence of Hall resistivity  $\rho_{xy}$ of the two samples,
filled symbols for $(La_{0.87}Sr_{0.13})OFeAs$, open symbols for
$La(O_{0.9}F_{0.1})FeAs$. One can see that the Hall resistivity is
positive for the sample $(La_{0.87}Sr_{0.13})OFeAs$ in wide
temperature regime, which is in sharp contrast with that of
$La(O_{0.9}F_0.1)FeAs$. (b) The temperature dependence of the Hall
coefficient $R_H$ taking in the zero field approach for the two
samples, filled symbols for $(La_{0.87}Sr_{0.13})OFeAs$, open
symbols for $La(O_{0.9}F_{0.1})FeAs$. This result clearly indicates
that our present sample has hole-like charge carriers for the
electron conduction in wide temperature region. But a much stronger
temperature dependence was observed which may suggest a stronger
multiband effect in the hole doped samples. } \label{fig.4}
\end{figure}

In order to know whether the Sr doped samples are really in the hole
doped regime, we measured the Hall effect for all the samples. As an
example, in Fig. 4(a) we show the Hall resistivity $\rho_{xy}$ for
both $(La_{0.87}Sr_{0.13})OFeAs$ and the electron doped sample
$La(O_{0.9}F_{0.1})FeAs$. It is clear that $\rho_{xy}$ is positive
at all temperatures below 200 K for $(La_{0.87}Sr_{0.13})OFeAs$
leading to a positive Hall coefficients $R_H=\rho_{xy}/H$. This is
in sharp contrast with the data of the electron doped sample
$La(O_{0.9}F_{0.1})FeAs$ \cite{ZhuXY}. The positive Hall resistivity
appears for all Sr-doped samples. In Fig. 4(b) the temperature
dependent Hall coefficient of the two samples are shown. One can see
that the Hall coefficient for the hole doped sample has much
stronger temperature dependence, which may suggest a stronger
multiband effect in the present sample. In addition, beyond about
200 K, the Hall coefficient drops to zero and becomes even slightly
negative. This change is just corresponding to the appearance of the
huge bump on the resistivity curve in the same temperature region.
Interestingly, in the electron doped samples, the Hall coefficient
also drops down when a little saturation of resistivity
occurs.\cite{ZhuXY}. This similarity may indicate an intimate
connection between these two different samples. It is clear that, in
wide temperature region, the Hall coefficient of the two samples has
different signs. The magnitudes of $R_H$ for the two samples at
about 100 K are in the same scale. If using the single band equation
$n=1/R_He$ to evaluate the charge carrier density, at 100 K, we
obtained $n(electron-doped)= 9.8\times10^{20}/cm^3$, while
$n(hole-doped)= 4.57\times10^{20}/cm^3$, both have a low charge
carrier density. This may give support to a theoretical proposal
that the iron based superconductor has very low superfluid
density.\cite{Singh} The positive sign of Hall coefficient in our
present Sr doped samples convinces us that they are indeed hole
doped. We also tried to substitute La with Ca at a concentration of
0.1, the result is the same as that reported by Kamihara et
al.\cite{Kamihara1}, that is no superconductivity was found.
Therefore it leaves a very interesting argument that the
superconductivity is not only controlled by the property of the
$FeAs$ layer, but also strongly influenced by the $LaO$ layer with a
subtle change.

\begin{figure}
\onefigure[width=8cm]{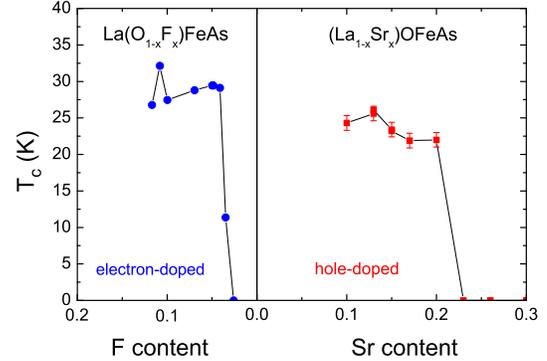} \caption{The generic phase diagram
depicted based on the data of our present system
$(La_{1-x}Sr_x)OFeAs$ and that of electron doped system
$La(O_{1-x}F_x)FeAs$. The phase diagram looks very similar to that
of cuprate superconductors.} \label{fig.5}
\end{figure}

Finally, in Fig.5, we depict a generic phase diagram by combining
our data from the hole doped system $(La_{1-x}Sr_x)OFeAs$ and the
data from the electron doped system $La(O_{1-x}F_x)FeAs$ of Kamihara
et al.\cite{Kamihara1}. Besides the somewhat flattened doping
dependence of $T_c$ in the intermediate doping regime,
interestingly, the phase diagram looks very similar to that of the
cuprates, which may give important clues to the mechanism of cuprate
superconductors. At this moment, we don't know whether there is also
a pseudogap in the normal state of the present iron-based
superconductors as appearing in underdoped cuprates\cite{Timusk}. A
detailed investigation on the properties of the hole doped samples
at different doping levels in present system is highly desired. The
similarity between the phase diagrams of the cuprates and the
iron-based system may imply that the superconductivity is in the
vicinity of some magnetic correlations, such as an antiferromagnetic
correlations/fluctuations in the cuprates, and ferromagnetic
correlations/fluctuations in the present iron-based materials. This
argument can get a support if the magnetic signal with a weak
ferromagnetic feature in the normal state is intrinsic to the
$LaOFeAs$ system. Regarding the superconductivity found in the hole
doped side, and combining the density of states calculated by
Dynamical Mean Field Theory (DMFT)\cite{Kotliar}, we suggest that
the $3d_{xy}$ orbit may paly an important role here. The new phase
diagram for this iron-based superconducting system without copper
may open a new era for the research of fundamental mechanism of
superconductivity, which will probably promote the solution to the
mechanism of cuprate superconductors. Our discovery of
superconductivity in the hole doped side will widely open the
territory for exploring new superconductors, hopefully leading to a
much higher superconducting transition temperature.

\section{Conclusion}

In summary, by substituting the tri-valence element La partially by
di-valence element Sr in LaOFeAs, we introduced holes into the
system and found superconductivity. The maximum transition
temperature is about 25 K at a doping level of x = 0.13. The
resistive onset transition temperature is rather stable in wide
doping region from 0.10 to 0.20, but no superconductivity was
observed beyond x = 0.23. Evidence for hole-like charge carriers has
been illustrated by Hall effect measurements. The general phase
diagram looks very similar to that of the cuprate, implying a very
fundamental constraint on the mechanisms of the two systems.

\acknowledgments

This work was supported by the Natural Science Foundation of China,
the Ministry of Science and Technology of China (973 Projects No.
2006CB601000, No. 2006CB921802), and Chinese Academy of Sciences
(Project ITSNEM).

Author Contributions: HHW designed and coordinated the whole
experiment, including the details about the doping and the
synthesizing, and did part experiments, analyzed the data and wrote
the paper. GM, LF and XYZ equally contributed to the synthesizing
and part of the magnetic measurements, LF did the structure
analysis. HY did the resistive and Hall measurements.

\end{document}